\documentclass[showkeys,superscriptaddress]{revtex4}
\usepackage[dvips]{graphicx}
\newcommand{\Na}{${\textrm {Na}}^+$}
\newcommand{\K}{${\textrm {K}}^+$}

\DeclareMathVersion{euler}
\SetSymbolFont{letters}{euler}{U}{eur}{m}{n}
\newcommand{\micro}{\mathversion{euler}$\mu$\mathversion{normal}}
\newcommand{\um}{{\micro}m}

\newcommand{\uF}{{\micro}F}

\newcommand{\uFcmsq}{\uF/$\textrm{cm}^2$}

\newcommand{\mScmsq}{{mS}/$\textrm{cm}^2$}

\newcommand{\GNabar}{\bar{G}_\textrm{{\tiny Na}}}

\newcommand{\GKbar}{\bar{G}_\textrm{{\tiny K}}}

\newcommand{\GL}{G_\textit{{\tiny L}}}

\newcommand{\ENa}{E_\textrm{{\tiny Na}}}
\newcommand{\EK}{E_\textrm{{\tiny K}}}

\newcommand{\EL}{E_\textit{{\tiny L}}}

\newcommand{\s}[1]{ \left[ {#1} \right] }
\newcommand{\p}[1]{ \left( {#1} \right) }

\begin{document}
\title{Josephson junction simulation of neurons}

\author{Patrick Crotty}\affiliation{Physics Department, Colgate University, Hamilton, NY, USA}
\author{Dan Schult}\affiliation{Mathematics Department, Colgate University, Hamilton, NY, USA}
\author{Ken Segall}\affiliation{Physics Department, Colgate University, Hamilton, NY, USA}


\begin{abstract}
With the goal of understanding the intricate behavior and dynamics
of collections of neurons,
we present superconducting circuits containing Josephson junctions that model 
biologically realistic neurons.  These ``Josephson junction neurons'' reproduce many
characteristic behaviors of biological neurons such as action potentials, refractory 
periods, and firing thresholds.  They can be coupled together in ways that mimic
electrical and chemical synapses.  Using existing fabrication technologies, large 
interconnected networks of Josephson junction neurons would operate fully in parallel.  
They would be orders of magnitude faster than both traditional 
computer simulations and biological neural networks.  Josephson junction neurons 
provide a new tool for exploring long-term large-scale dynamics for networks of neurons.
\end{abstract}

\keywords{neuron model | neural networks | Josephson junction | Hodgkin-Huxley } 

\maketitle

How do large networks of neurons organize, communicate and collaborate to create the 
intrinsic behaviors and dynamics of the brain?  Over the past century, individual 
neurons have been studied at the cellular, compartmental and molecular level.  
Synaptic models, while still somewhat rudimentary, accurately reflect many basic 
features of synapses and how they modify signals between neurons.  Today, it is 
becoming feasible to explore networks of neurons behaving as units, how they 
synchronize, provide top-down or bottom-up feedback, and encode sensory information.  
This exploration is an important step toward understanding the brain, which will 
require multi-scale analysis with models of collective behavior at many different 
levels simultaneously.

As part of this effort it is important that we understand how networks of neurons behave
on the scale of thousands to tens of thousands of neurons, the size of a typical neocortical 
column.  Large scale digital simulation projects such as the Blue Brain 
\cite{markram2006,king2009component} and PetaVision 
\cite{petavision} 
projects have demonstrated that the limitations of inherently serial computer processors
can be improved by effective parallel computing designs.  But simulation time remains a 
significant hurdle to including biologically realistic features in large scale simulations.
Analog simulations using VLSI circuitry to mimic neurons and synapses are improving in
realism and speed, but they are still limited by complexity and power consumption.
We propose a new direction for analog simulation of large scale networks of biologically 
realistic neurons.  Using superconducting Josephson junctions to model
neurons connected with real-time synaptic circuitry, we can explore
neural network dynamics orders of magnitude faster than current digital or
analog techniques allow.  Using these circuits, we hope to learn about neural
interactions such as synchronization, long term dynamics and bifurcations, feature 
identification and information processing.  The long term goal is to understand group
behavior of neurons sufficiently to use them as building blocks for studying 
larger scale neural networks and brain behavior.

\begin{figure}
 \begin{center}
   \includegraphics[width=8.4cm]{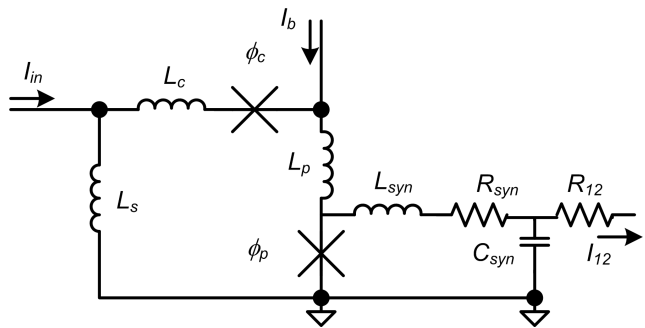}
     \caption{ \label{fig:circuit}
     Circuit diagram for the JJ Neuron (left loop) connected 
     to a model chemical synapse (right loop).  In general, many
     synapses could connect to a single JJ Neuron.  Orientation of junction
     phases is chosen for clockwise current in the left loop.
     }
 \end{center}
\end{figure}

Our basic circuit unit (the JJ Neuron) involves two Josephson junctions connected in 
a loop as shown on the left side of Fig.~\ref{fig:circuit}.  
The individual junctions behave phenomenologically like ion channels:  one 
corresponds to a depolarizing current (like \Na), and the other to a hyperpolarizing 
current (like \K).
Enhancements are possible, of course, and the inclusion of a third junction
could allow for behaviors such as bursting that generally require at least
three currents.
The circuit displays many features of biologically realistic neurons such as 
the evocation of action potentials (firing) in response to input currents or pulses,
input strength thresholds below which no action potential is evoked, and refractory periods 
after firing during which it is difficult to initiate another action potential
\cite{izhikevich2004which, hoppensteadt1997}.  

The JJ Neuron is a variation of well developed Rapid Single Flux Quantum (RSFQ) 
\cite{likharev1991,bunyk2001rsfq,brock2001rsfq} circuitry and is thus straightforward to fabricate.  
RSFQ circuits using $20,000$ junctions have been fabricated \cite{brock2000superconductor}, 
so we estimate that a single chip 
could model as many as $N=10,000$ neurons, about the number of neurons in a cortical column.  
For larger brain regions, chips could be connected together.  A simulated action 
potential takes about 50 picoseconds, and all neurons act in parallel.  
Fig~\ref{tab:speed} shows a comparison of the speed of a JJ neuron with that 
of biological neurons and digital simulations using established models.  
The figure of merit displayed is the number of action potentials per neuron per 
second.  Speed depends on the arithmetic complexity of the model employed, 
the number $N$ of neurons simulated and their connectivity.  
We show speeds for sparse (no connections between neurons) and dense 
(all neurons connected to all others) networks as two extremes.  
While our estimates of speed are crude, they suggest that JJ Neurons are several 
orders of magnitude faster than either digital simulations or biological systems.  
This speed is due to three advantages:  
a) Josephson junctions have a shorter 
characteristic time than the conventional transistors used in computer processors;  
b) the JJ Neuron takes fewer cycles 
to simulate an action potential than digital simulations; 
and c) JJ Neurons compute in parallel using analog connections 
that do not affect computation time.

\begin{figure*}
  \begin{center}
    \newcommand{\spaceline}{\rule{0pt}{2.6ex} \\ \hline }
    \begin{tabular}{|r|c|c|c|c|}
    \hline
   &  &\multicolumn{3}{|c|}{{\bf (AP/neuron/s)}} \\\cline{3-5}
  {\bf Model} \hspace{1cm} & {\bf FLOPS/AP} & {\bf{\em N} = 1} & {\bf {\em N}=1000 (sparse)} & {\bf {\em N}=1000 (dense)} \\
   \hline 
  Integrate and Fire & 5 & 2.0$\times10^8$ & 2.0$\times10^5$ & 5.0$\times10^2$ \spaceline
  Fitzhugh-Nagumo   & 72 & 1.4$\times10^7$ & 1.4$\times10^4$ & 4.8$\times10^2$ \spaceline
  Izhikevich        & 13 & 7.7$\times10^7$ & 7.7$\times10^4$ & 5.0$\times10^2$ \spaceline
  Hindmarsh-Rose   & 120 & 8.3$\times10^6$ & 8.3$\times10^3$ & 4.7$\times10^2$ \spaceline
  Hodgkin-Huxley  & 1200 & 8.3$\times10^5$ & 8.3$\times10^2$ & 3.1$\times10^2$ \spaceline
  Mammalian CNS    & --- & 1.0$\times10^3$ & 1.0$\times10^3$ & 1.0$\times10^3$ \spaceline
  JJ Neuron        & --- & 2.0$\times10^{10}$&  2.0$\times10^{10}$ & 2.0$\times10^{10}$ \spaceline 
    \end{tabular}
    \caption{ \label{tab:speed}
    Approximate number of action potentials(APs) per 
    neuron per second for digital simulations, mammalian 
    experiments and the JJ Neuron.
    FLOPs/AP is an estimate of the floating point operations 
    required for one AP in each digital model \cite{izhikevich2004which}, 
    and we assume a conservative 2 FLOPs per connection. 
    We assume a CPU speed of $10^9$ FLOPs/second for the digital simulations. 
    The right three columns describe the speed of a network with $N$ neurons. 
    Sparse and dense connectivity represent extreme estimates of 
    the computational time due to connections between neurons. 
    Connections have no speed impact on experimental
    and JJ Neuron models as they are naturally parallel.}
  \end{center}
\end{figure*}

This simple circuit simulates a single-compartment (space clamped) neuron and
follows the spirit of mathematical neuron models such as 
Hodgkin and Huxley \cite{hodgkin1952quantitative}, 
Fitzhugh-Nagumo \cite{fitzhugh1961,nagumo1962}, 
Morris-Lecar \cite{morris1981}, 
Hindmarsh-Rose \cite{hindmarsh1982} 
and Izhikevich \cite{izhikevich2004which}.   
Each of these models describes the membrane potential (voltage) of the neuron as it
varies in time.  They also model, to varying degrees of detail, the ion channels 
which control the voltage 
via the flow of ions across the membrane.  The ion channels open and close depending 
on the voltage thus providing the nonlinear feedback characteristic of neurons.
Models differ in how many ion channels are described and their dynamic complexity.  
As with many of these mathematical models, our intent is not to model ion channel 
dynamics or chemical reactions in detail but rather to provide basic biological 
realism which allows us to explore the impact of network topology and connectivity 
strength.

The JJ Neuron attempts to mimic important neuronal behaviors:
action potentials, firing thresholds and refractory periods.
The primary identifying feature of a neuron is the action potential (AP), i.e.
a spike or pulse in voltage across the neuron membrane as the neuron ``fires''.
AP firing rates and inter-spike interval distributions are used to encode information in the brain 
\cite{rieke1999}.  
APs require at least two voltage-gated ionic currents, generally \Na and \K.  
The inward \Na current produces the rising phase
of the pulse, while the outward \K current restores the membrane to its 
resting potential \cite{johnston1995}.  These currents have different time scales,
resulting in the characteristic shape of the action potential.  
The dynamics of biological voltage-gated ionic currents
are complex, and subsequent models have substantially expanded on those originally
proposed by Hodgkin and Huxley \cite{vandenberg1991,clay2005}.  
The JJ Neuron, like most neuron models, simplifies the voltage dependence of ionic currents,
modeling them by junctions whose dynamics are governed by familiar second order 
differential equations.  It does include two ionic channels and supports APs.
Other important features of a neuron model include the firing threshold and refractory period.
The firing threshold is a level of external stimulus below which
the response is negligible, while above it an AP is triggered.  
The refractory period is the period of time after triggering an AP during which it is 
difficult to evoke a second one.
Biophysically, the refractory period is determined by the time it takes the 
ion channel proteins to ``reset'' to their initial configurations.

\section{Single neuron model}
Josephson junctions consist of two superconductors separated by a 
thin insulating barrier \cite{josephson1962possible}.  Electrons in each 
superconductor are described by a coherent wave-function with a definite phase.  
The phase difference from one side of the barrier to the other is the 
so-called Josephson phase $\phi$, which controls all of the electrical 
properties of the junction, including the junction's voltage and 
current.  Current can flow through the device without creating a voltage.
This so-called super-current can be increased to the junction's critical
current $I_0$ above which a voltage develops given by $V=(\Phi_0/2\pi)d\phi/dt$, 
where $t$ is time, $\Phi_0 = h/e$ is the flux quantum, $h$ is Planck's constant and $e$ is the 
electron charge.  Normalizing current to this critical current, the current
through a junction $i$ depends on the phase $\phi$ through the relation
\cite{orlando1991foundations}:
\begin{equation}
  \label{eqn:pendulum}
   i = \ddot{\phi} + \Gamma\dot{\phi} + \sin(\phi). 
\end{equation}
The dot notation refers to differentiation with respect to normalized time 
$\tau$, where $\tau^2 = t^2\Phi_0C/2\pi I_0$, with $C$ the capacitance of the 
junction. The normalized damping parameter is $\Gamma^2 = \Phi_0/2\pi I_0R^2C$, 
where $R$ is the resistance of the junction.  
The critical current, conductance ($1/R$) and capacitance of 
a junction are assumed to scale linearly with the cross-sectional area ($A$) of the junction.

Equation \ref{eqn:pendulum} displays a useful analogy between the dynamics of a junction
and that of a pendulum, allowing us to describe the action potential-like 
pulse generated by the circuit.  
Equation \ref{eqn:pendulum} is exactly the equation of motion for a damped and 
driven pendulum, where $\phi$ represents the angle of deflection, $\Gamma$ is the 
normalized drag constant, and $i$ represents a driving torque.  
If $i$ is constant in time, long term solutions include both static tilting 
(gravity balancing the torque)
and whirling modes (where the torque overcomes gravitational forces).  
A third motion is also possible that combines these two.  
For appropriate parameters, and time dependent torque $i$, the pendulum will whirl
over just once and settle back to a tilted state.
In a Josephson junction, the phase whirling over just once creates a 
magnetic flux pulse, called a Single-Flux-Quantum pulse (SFQ pulse)\cite{likharev1991}.  
This pulse has a similar shape and area each time it is stimulated.  
Pulses of this type form the action potentials in our neuron model.

The JJ Neuron circuit shown in Fig.~\ref{fig:circuit} connects two Josephson 
junctions in a superconducting loop.  
This basic structure is a simplified superconducting digital 
component from RSFQ logic circuitry called a ``DC-to-SFQ converter'' \cite{bunyk2001rsfq}.
We call the two junctions the pulse 
junction and the control junction, denoted by subscripts $p$ and $c$ respectively.
Two incoming currents (normalized to $I_{0p}$) are called the input current $i_{in}$ and the bias
current $i_b$ which provides energy to the circuit.  
The signal to the synapse is
the voltage across the pulse junction $v_p = \dot{\phi_p}$, 
though the magnetic flux through the loop could be used for an 
inductive synapse circuit.
The figure shows a model synaptic circuit connected above the pulse junction and thus driven by $v_p$.
The circuit parameters are the indicated branch inductances $L_s$, $L_p$, and $L_c$ 
which we scale by their sum $L_{total}$ to obtain $\Lambda_s$, $\Lambda_p$ and 
$\Lambda_c$.  
Using current conservation and fluxoid quantization \cite{orlando1991foundations}, we obtain 
two equations of motion:
\begin{equation}
  \label{eqn:pulse}
  \ddot{\phi_p} + \Gamma\dot{\phi_p} + \sin(\phi_p) = 
  i_p = - \lambda (\phi_c + \phi_p) + \Lambda_s i_{in} + (1-\Lambda_p) i_b,
\end{equation}
\begin{equation}
  \label{eqn:control}
  \eta \s{\ddot{\phi_c} + \Gamma\dot{\phi_c} + \sin(\phi_c)} = 
  i_c = - \lambda (\phi_c + \phi_p) + \Lambda_s i_{in} - \Lambda_p i_b.
\end{equation}
with coupling parameter $\lambda = \Phi_0/2\pi L_{total} I_{0p}$, and geometric 
parameter $\eta=A_c/A_p$.
To make the comparison with RSFQ circuitry more clear, DC-to-SFQ converters usually 
have $\Lambda_s\approx 1$, $\Lambda_p=\Lambda_c=0$ and $\eta<1$, 
whereas typical parameters for a JJ Neuron are $\Lambda_c=0$, $\Lambda_s=\Lambda_p=0.5$,
 and $\eta=1$.

To explain how the circuit provides an action potential and then recharges to become ready
to provide another, we describe the mechanism for generating a typical pulse.
In equilibrium with $i_{in}=0$, the bias current splits between pulse and control 
junctions and is large enough to put each junction just below its critical 
current.  They are primed to achieve a whirling state. 
For $i_{in}>0$, the input current acts to push the control junction away from 
whirling while pushing the pulse junction toward the whirling state.
When the input current exceeds a threshold, it initiates the action potential:  
A voltage appears across the pulse junction creating magnetic flux in the loop which
is analogous to a neuron's membrane potential rising.
The flux in turn induces current in the loop pushing the control junction closer to 
its whirling state.  As the flux builds, the control junction starts to whirl, 
draining the flux in the loop, ensuring that the pulse junction stops whirling and 
restoring the system so it can fire again.
This process can repeat so long as the incoming current is held above threshold levels.
The time lag between the pulse junction whirling, flux building and control junction
responding to that flux is what creates the refractory period during which it is
extremely difficult to initiate another pulse.
The analogy goes beyond the membrane potential's relation to the flux
$\Phi = \lambda(\phi_p+\phi_c)$ normalized by $L_{total}I_{0p}$.  
The pulse voltage $v_p$ is analogous to a polarizing ionic current such as \Na
while the control voltage $v_c$ is analogous to a hyperpolarizing 
ionic current such as \K.  

The flux $\Phi$ in the JJ Neuron corresponds to the neuron membrane potential $V_m$.  
Voltages across the pulse junction, $v_p$, and control junction, $v_c$, correspond to 
the \Na current $I_{Na}$ and \K current $I_K$ respectively.  
The input current $i_{in}$ corresponds to incoming post-synaptic 
current $I_{syn}$.  With these correspondences, the equation relating 
change in membrane potential to ionic currents can be written for the JJ Neuron
\begin{equation}
  {1\over\lambda} \dot{\Phi} = v_p + v_c.
  \label{eqn:fluxcable}
\end{equation}
The analogous equation for a neuron is the Cable Equation 
for a space-clamped 2-channel neuron\cite{johnston1995}:
\begin{equation}
  C\dot{V}_m=I_K+I_{Na}+I_{syn}.
  \label{eqn:cable}
\end{equation}
The equations differ in that the synaptic current in (\ref{eqn:cable}), $I_{syn}$,
does not appear analogously as $i_{in}$ in (\ref{eqn:fluxcable}).  Synaptic current 
does, however, appear implicitly since the phases $\phi_p$ and $\phi_c$ depend on $i_{in}$
through (\ref{eqn:pulse})-(\ref{eqn:control}).  
The results of simulations below suggest that
these terms have qualitatively similar effects.  
Like biological APs, the JJ Neuron AP is produced by the 
interaction of activating and restoring processes.  

\section{Single neuron characteristics}
\begin{figure}
 \begin{center}
   \includegraphics[width=8.4cm]{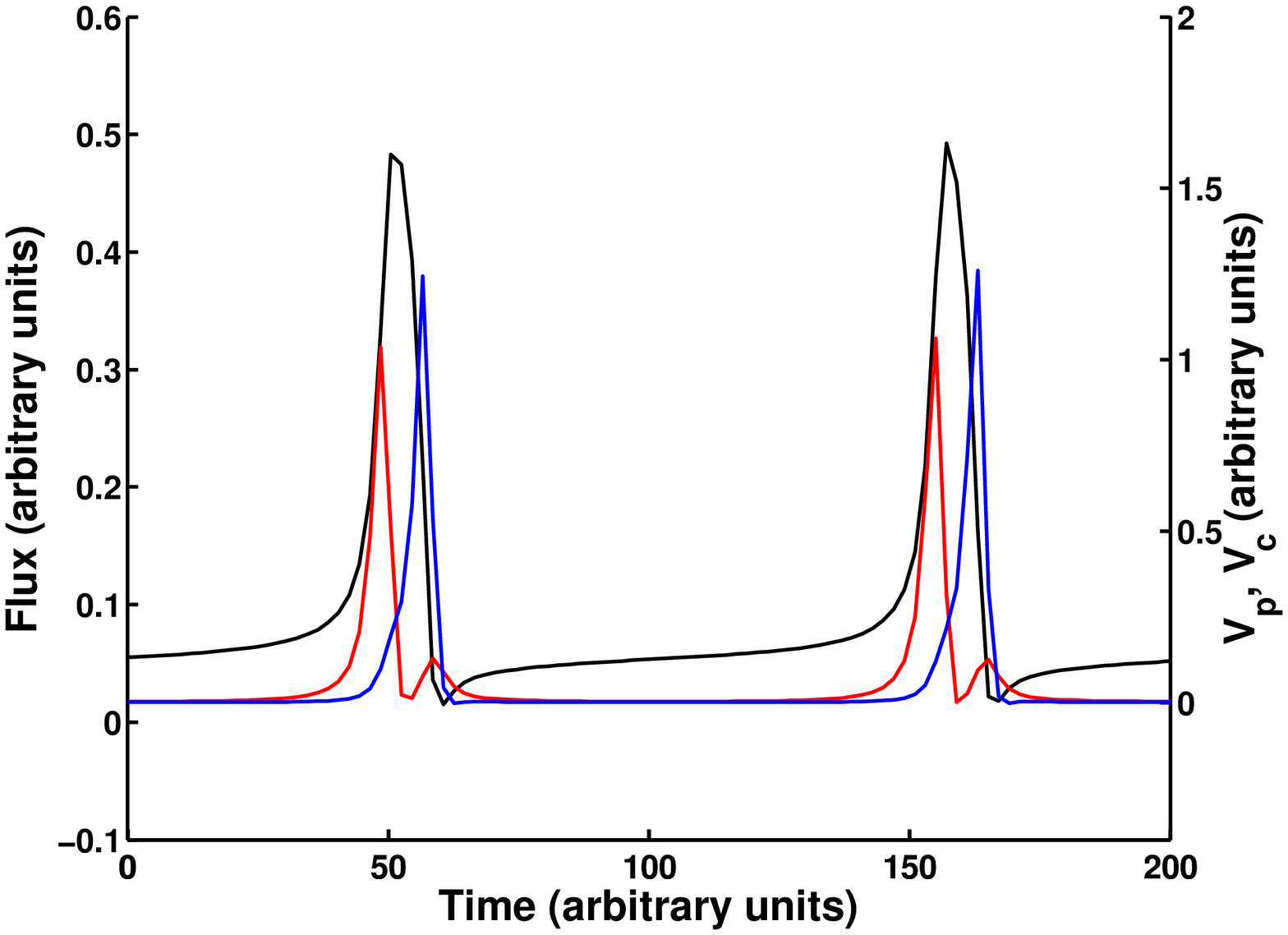}
   
   \includegraphics[width=8.4cm]{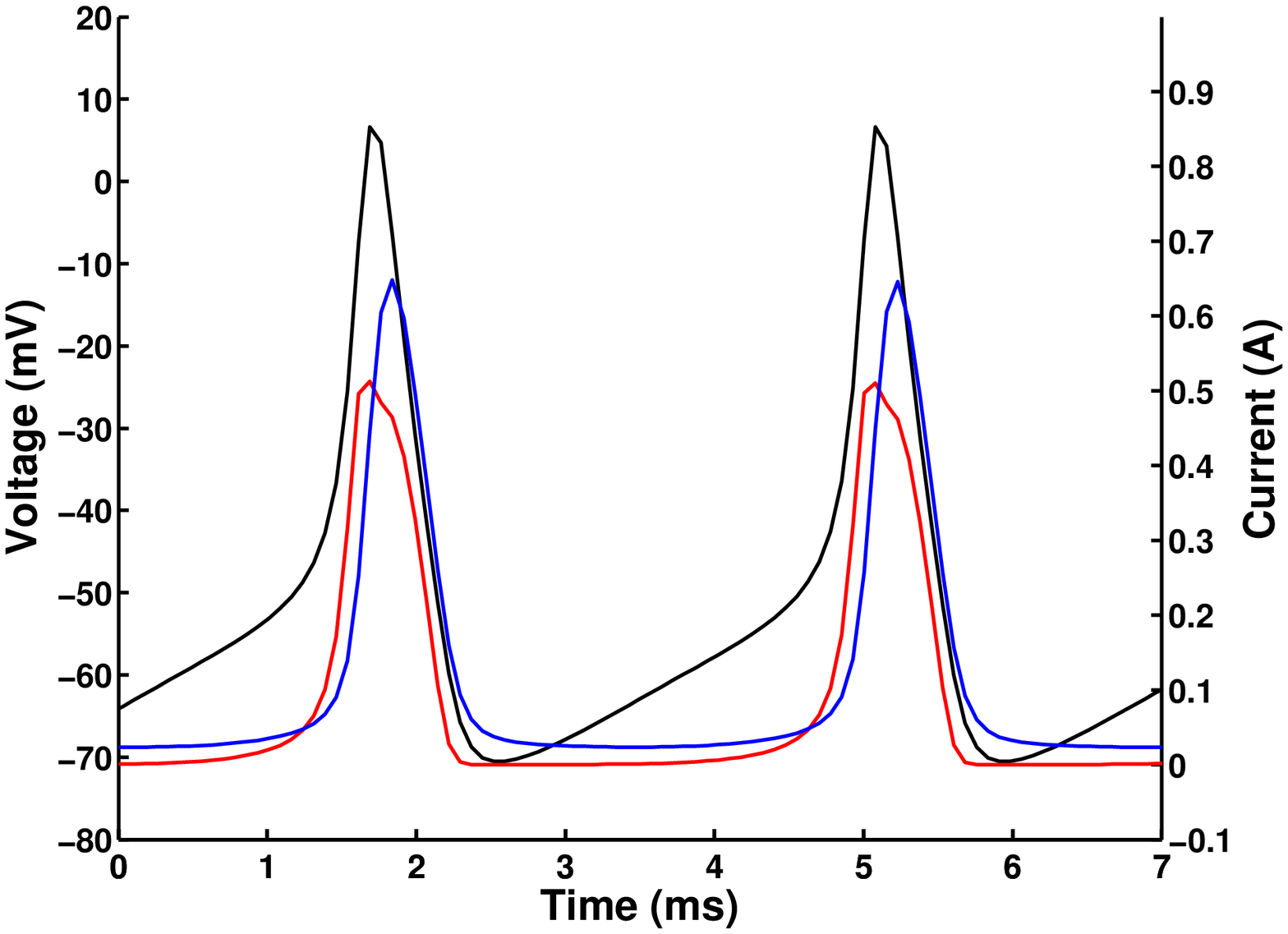}
   \caption{ \label{fig:AP}
     A. Time profile for action potentials in the JJ Neuron model.
     Parameter values for JJ Neuron calculations are 
     $\lambda=0.1$, $\Gamma=1.5$, $\Lambda_s=\Lambda_b=0.5$, $\eta=1$. 
     Input DC current is $i_{in}=0.21$.
     B. Time profile for action potentials in the Hodgkin-Huxley model.
     The flux (black) in the JJ Neuron corresponds to the membrane 
     potential (black) in the Hodgkin-Huxley model.  
     Similarly, the voltages $v_p$ (red) and $-v_c$ (blue) in the JJ Neuron model correspond to
     the currents $-I_{Na}$ (red) and $I_K$ (blue) in Hodgkin-Huxley model.
     Parameter values for all Hodgkin-Huxley calculations are 
$C_m = 1.01$ \uFcmsq; $\GNabar = 120$ \mScmsq; $\GKbar = 36$ \mScmsq;
$\GL = 0.3$ \mScmsq; $\ENa$ = 50 mV; $\EK$ = -77 mV; $\EL = -54.4 mV$; and
$T = 18.5 ^\textrm{o}\textrm{C}$ unless stated otherwise.  In this simulation
an external stimulating current of $236 nA$ is applied.
The Hodgkin-Huxley neurons are modeled as isopotential cylinders with
lengths and diameters of 500 \um.
   }
 \end{center}
\end{figure}

Single-neuron characteristics include the action potential,
firing threshold and refractory period.
JJ Neurons reproduce all three of these behaviors. 
Fig.~\ref{fig:AP} shows the 
voltage trace and ionic currents for APs in the JJ Neuron and
Hodgkin-Huxley models.  

\begin{figure}
   \begin{center}
     \includegraphics[width=8.4cm]{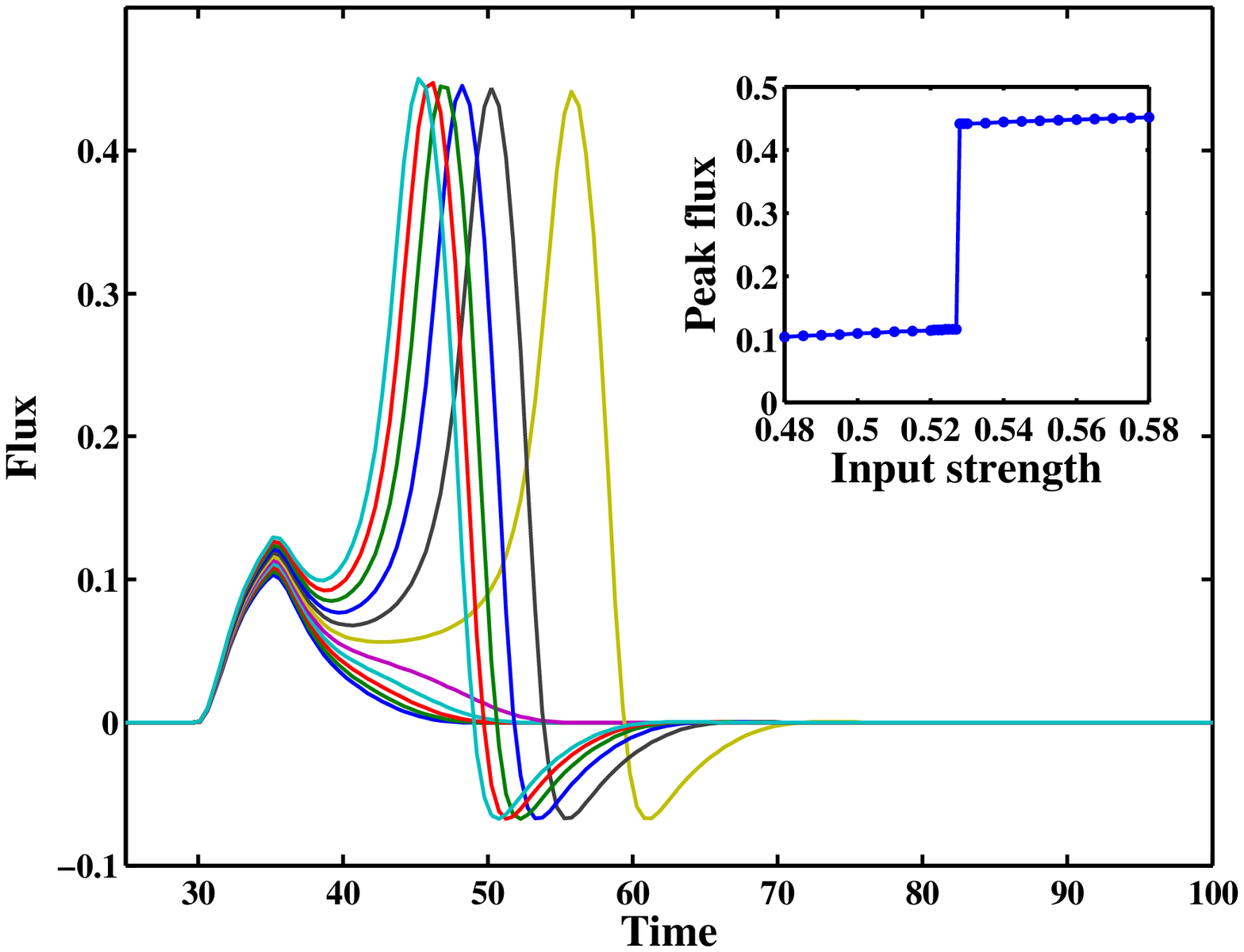}
     \caption{ \label{fig:threshold}
     The threshold response of flux time-traces colored by the strength of impulse input signals.
     Stronger inputs lead to action potentials.
     The inset graph shows how peak response depends on the strength 
     of the input pulse.  
     Parameter values as for Fig.~\ref{fig:AP}A except $\Gamma=1.0$.
     Input current is a single square pulse of width 5 and varied height initiated at $t=30$.
     }
   \end{center}
\end{figure}

Fig.~\ref{fig:threshold} demonstrates the firing threshold.
It shows the response to a brief current pulse of increasing 
strength along with an inset showing the threshold dependence of voltage peak on stimulus 
strength.  Stimuli below threshold evoke a small (sub-threshold) response, while
above the threshold full APs occur.

\begin{figure}[b]
   \begin{center}
     \includegraphics[width=8.4cm]{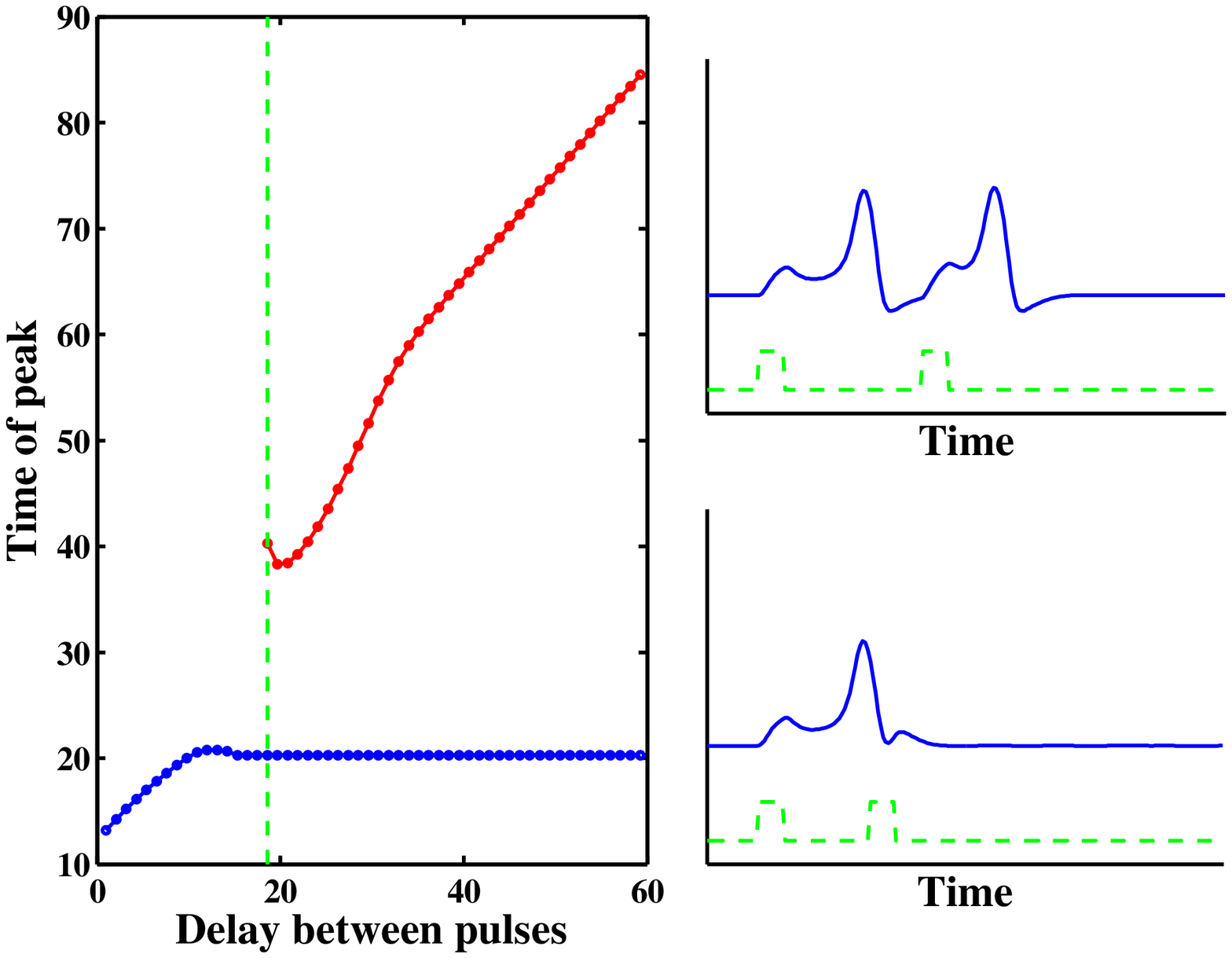}
     \caption{ \label{fig:refractory}
     a) Response to twin pulse inputs with delay shows the 
     absolute refractory period.  On the left, the time of 
     the first (blue) and second (red) response
     peak is graphed against the delay between input pulses.  
     The vertical
     line represents the delay below which no second pulse is created.
     On the right, time profiles are shown for input currents (green) 
     and output voltages (blue).
     In b) two pulses are generated while in 
     c), the delay is too small and no second pulse arises.
     Parameter values as for Fig.~\ref{fig:AP}A except $\Gamma=1.0$.
     Input pulses are twin square pulses of height 0.54, width 5 and
     varied delay between pulses.  The first pulse initiates at $t=30$.
     }
   \end{center}
\end{figure}

Fig.~\ref{fig:refractory} demonstrates the refractory period of the JJ Neuron.
It shows the peak response times to two brief 
current stimulus pulses as the time between pulses varies.
Current stimuli which are too closely spaced fail to produce a second AP.  
We emphasize that this 
refractory period is not explicitly hard-coded into the model (as for 
integrate-and-fire models \cite{gabbiani1998}), but appears naturally
as a result of differences in the time scales of the activating and 
restoring processes just as for biological neurons.
The slight upturn in the curve for the second peak corresponds to interference 
between two closely spaced action potentials. 
In parts b-c are shown time profiles of APs responding to two brief current 
pulses.  For large delay between pulses, the firing time of the second AP depends linearly on delay.
When the delay is sufficiently short, no second AP occurs.

\section{Connecting neurons} 
Synaptic models connect individual neuron models to form a network.
Synapses control communication, signal transfers, and timing \cite{kandel2000}.
They can operate by means of direct electrical connections (electrical
``gap junction'' synapses) or chemical intermediates called 
neurotransmitters (chemical synapse).  
Electrical synapses are modeled by direct connection between JJ Neurons.
We focus on chemical synapses here as they are both more complex and more common.
Chemical synapses can be excitatory or inhibitory, moving the 
post-synaptic neuron closer to or farther from threshold respectively.  
In chemical synapses, an AP causes the release of neurotransmitter molecules
which diffuse across the synaptic cleft, bind to receptors on the post-synaptic 
membrane, and induce input currents in the post-synaptic neuron.  
The net effect on AP transmission across the 
synapse is a delay in transmission and the spreading of the pulse.
We model the synaptic process using a resonant circuit 
attached to the pulse junction, shown on the right hand side of 
Fig.~\ref{fig:circuit}.  The output is taken across the capacitor and sent 
through a resistor to the input of a post-synaptic neuron.  
If the bias 
current applied to the JJ neuron is positive (negative) with respect to ground, 
then the synapse is excitatory (inhibitory).

The equations describing our synaptic circuit and the coupling to the 
next neuron downstream come from Kirchhoff's laws and involve 
normalized parameters describing the resonant frequency 
$\Omega_0=\tau/\sqrt{L_{syn}C_{syn}}$, 
quality factor $Q=\Omega_0R_{syn}C_{syn}/\tau$, 
synaptic coupling $\Lambda_{syn}=L_{syn}/L_{total}$, 
and resistive coupling to the next neuron, 
$r_{12}$, normalized by $R_{0p}$. 
We find that the voltage across the capacitor, $v_{out}$ (normalized by $\Gamma I_{0p}R0_{p}$), 
and the current coupled to the next neuron, $i_{12}$ (normalized by $I_{0p}$) are given by:
\begin{equation}
 {1\over \Omega_0^2}\ddot{v}_{out} + {Q\over \Omega_0}\dot{v}_{out} + v_{out} = v_p
 - {Q\Omega_0\Lambda_{syn}\over \lambda}i_{12} - {\Lambda_{syn}\over\lambda}\dot{i}_{12},
\end{equation}
\begin{equation}
  {\Lambda_{syn}\p{1-\Lambda_{syn}}\over \lambda}\dot{i}_{12} +{r_{12}\over\Gamma}i_{12}= 
  v_{out} - \Lambda_{syn}\p{v_{c2} + v_{p2}}.
\end{equation}
Here $v_{p2}$ and $v_{c2}$ represent the voltage across the pulse and control 
junction of the next neuron, respectively.

In addition, the synapse circuit's back-action on the JJ Neuron circuit creates 
two additional terms on the right side of (\ref{eqn:pulse}):
$-i_{12} - \lambda v_{out}/\p{\Lambda_{syn}\omega_0^2}$. 
For the simple demonstrations of coupling behavior shown here, this back-action was 
not strong enough to warrant buffer junctions, as is commonly needed 
with RSFQ circuits \cite{kadin1999introduction}.  However, for more complex circuits buffer 
junctions will most likely be needed.

\begin{figure}
   \begin{center}
     \includegraphics[width=8.4cm]{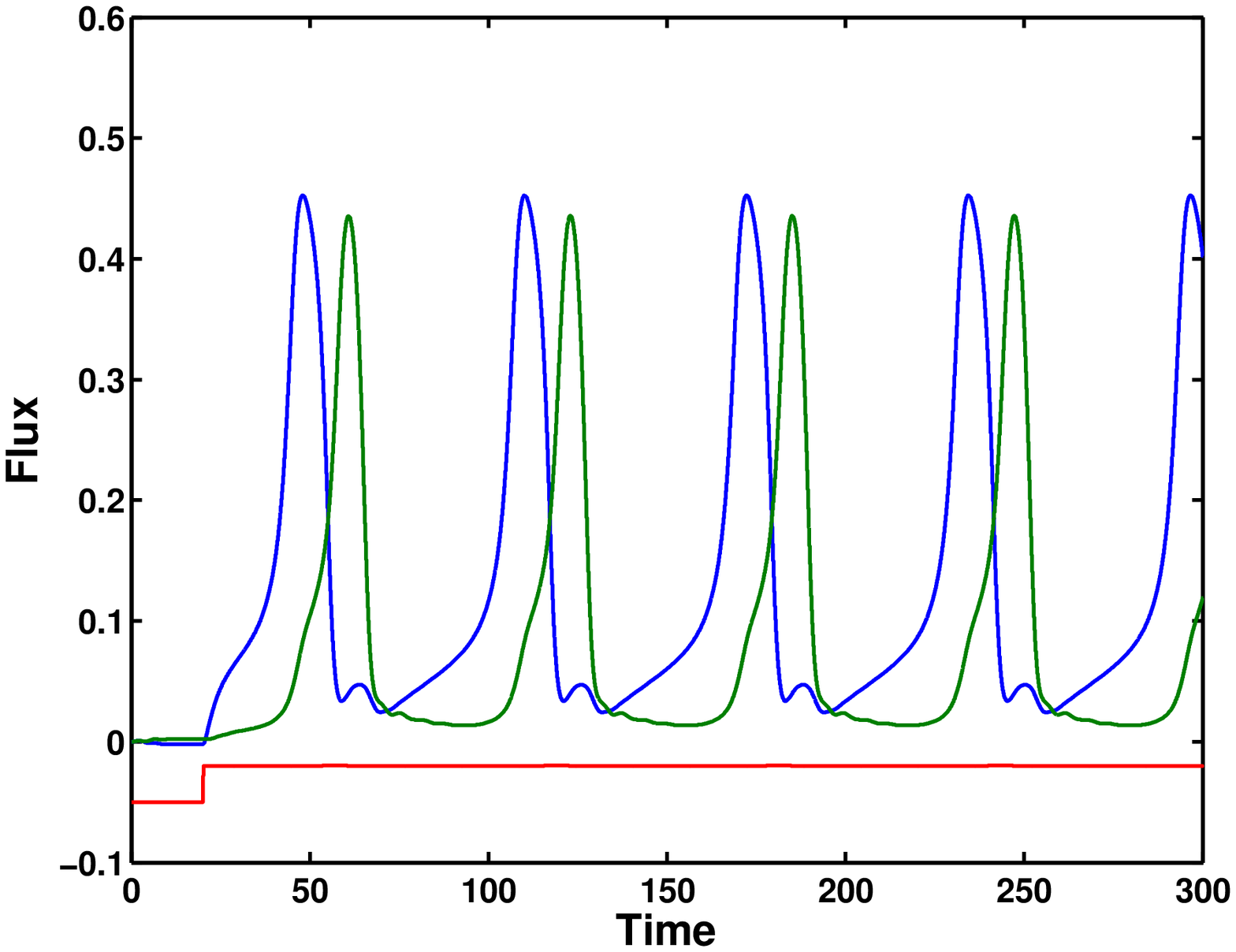}
     \caption{ \label{fig:coupled}
     Excitatory synaptic coupling.  
     One JJ neuron (blue) drives another (green).  
     The system is initially quiet.  At time 50, a constant 
     input current (red) causes the first neuron to fire repetitively.
     Excitatory coupling induces repetitive firing of the second neuron. 
     Parameter values as for Fig.~\ref{fig:AP}A except $i_b=2.0$, $\Gamma=2.0$.
     Synaptic parameters are $\Omega=1.0$, $Q=0.05$, $\Lambda_{syn}=0.3$, $r_{12}=1.4$.
     Input current to first neuron is a DC current of $i_{in}=0.3$ initiated at $t=20$.
     }
   \end{center}
\end{figure}

\begin{figure}
   \begin{center}
     \includegraphics[width=8.4cm]{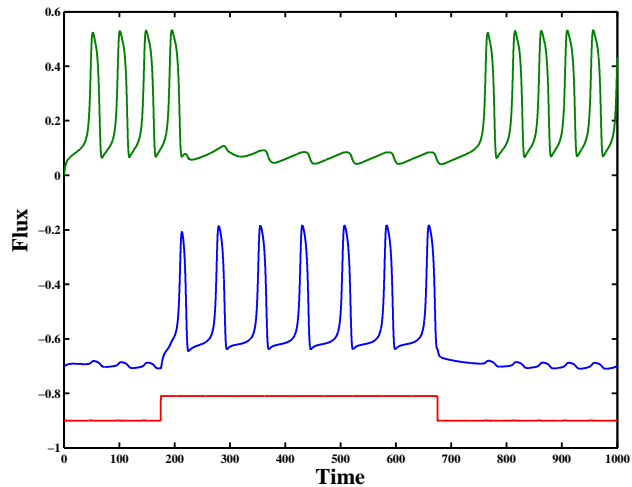}
     \caption{ \label{fig:inhib}
     Inhibitory synaptic coupling.  
     The second neuron (green) is configured to fire repetitively.  
     When the first neuron (blue) is activated by an external input current (red), 
     it inhibits the second.  Both return to their original 
     state after the external stimulus to the first neuron is removed.
     Parameter values as for Fig.~\ref{fig:coupled} except $i_{b1}=-1.9$, $i_{b2}=1.76$, 
     $\Lambda_{s2}=0.65$, and $\lambda_{p2}=0.35$.
     This pus the second neuron into a repetitive firing mode.
     Synaptic parameters are also the same except $r_{12}=0.6$.
     Input current to first neuron is a single square pulse of height 0.3, width 360, at $t=240$.
     }
   \end{center}
\end{figure}

We demonstrate synaptic behavior for inhibitory and excitatory coupling 
in a two-neuron setting.  
The first neuron is connected to the second via a synaptic connection as 
shown in Fig.~\ref{fig:circuit}.  
The results are shown in Fig.~\ref{fig:coupled}-\ref{fig:inhib}.  

\begin{figure}
   \begin{center}
     \includegraphics[width=8.4cm]{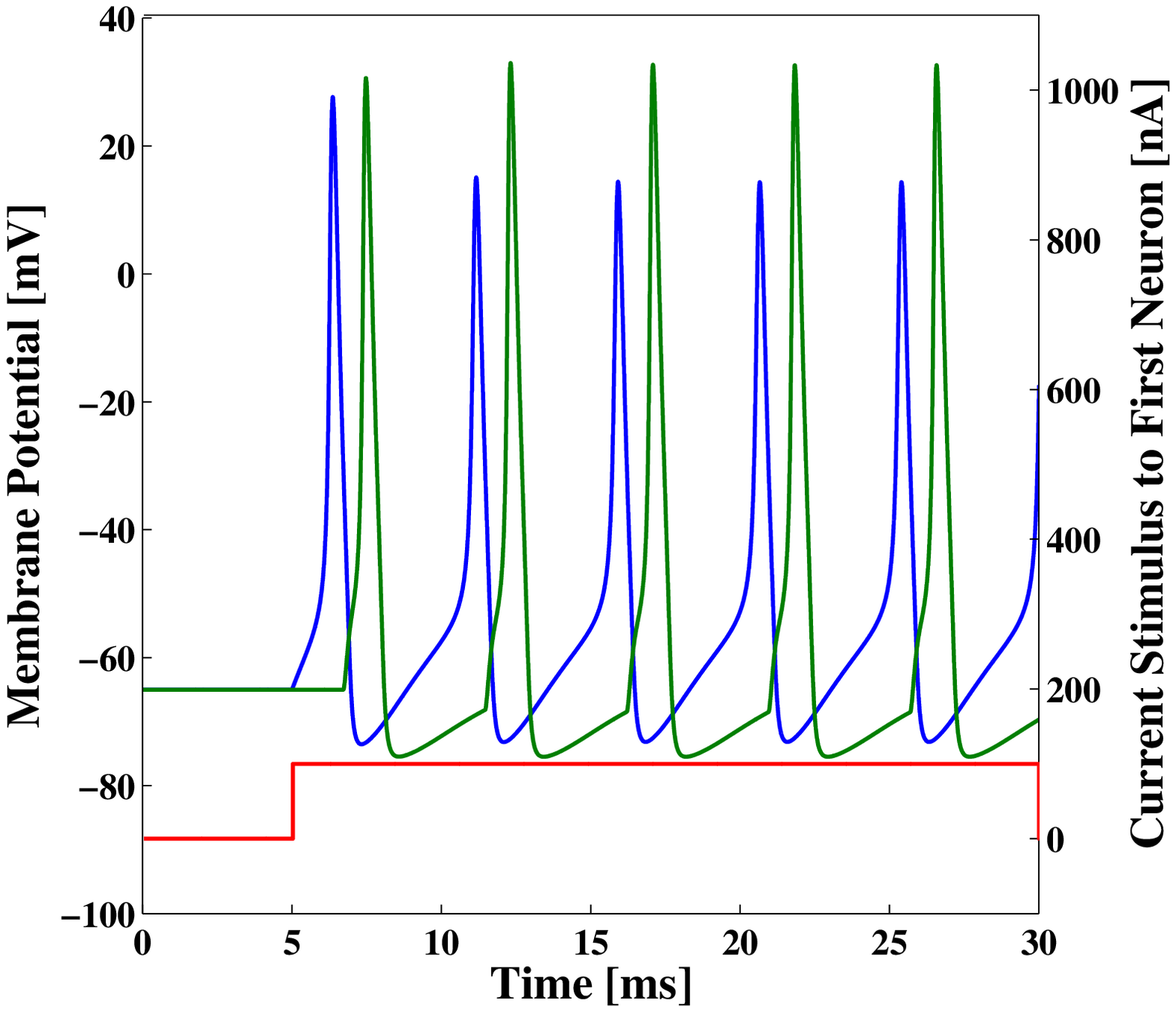}
     \caption{ \label{fig:HHcoupled}
     Behavior of two Hodgkin-Huxley model neurons coupled
     by a simple excitatory synapse model.  The synapse uses a positive ``alpha'' 
     function \cite{johnston1995} with maximum current 400 nA and
     time scale 0.1 ms which is triggered at -30 mV on the downward slope of the
     presynaptic neuron's action potential.  The presynaptic neuron (blue) is
     made to fire repetitively 
     by means of a constant current input (red curve) initiated at time 5 ms.  
     Excitatory coupling then causes repetitive firing of the postsynaptic
     neuron (green). 
     }
   \end{center}
\end{figure}

In Fig.~\ref{fig:coupled}, we show excitatory coupling from neuron 1 to neuron 2.  
Neuron 1 receives an external stimulus, 
and its output drives neuron 2, which responds by firing 
at the same rate but out of phase with neuron 1.  

\begin{figure}
   \begin{center}
     \includegraphics[width=8.4cm]{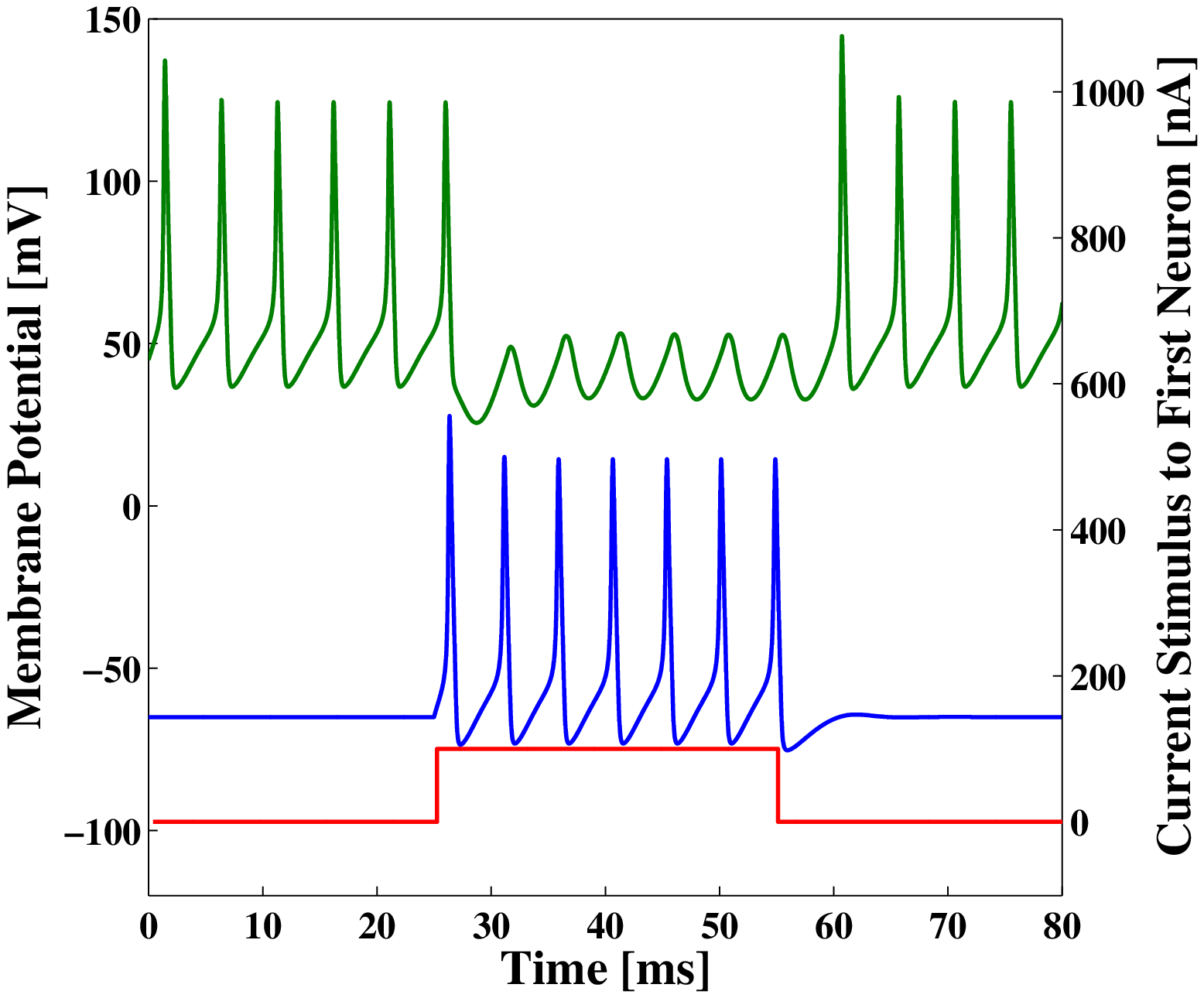}
     \caption{ \label{fig:HHinhib}
     Behavior of two Hodgkin-Huxley neurons coupled with an inhibitory synapse 
model using a negative alpha function with maximum current 215 nA and
time scale 1 ms.  The second neuron (green) has
$\EL$ set to -15 mV to make it fire repetitively in the absence of external inputs, 
while the first has $\EL$ set at the standard -54.4 mV value to make it normally
quiescent.  When the first neuron (blue) fires as the result of an external input 
current (red), it inhibits the second.  Both return to their original state after 
the external stimulus to the first neuron is removed.
     }
   \end{center}
\end{figure}

In Fig.~\ref{fig:inhib}, we show inhibitory coupling from neuron 1 to neuron 2.  
Neuron 2 is configured to fire repetitively by increasing its bias current.
When neuron 1 is not firing, this external stimulus causes neuron 2 to fire repeatedly.
However, when neuron 1 is stimulated, it inhibits the firing of neuron 2
even though neuron 2 continues to receive the external stimulus.  
Both return to their original state after the stimulus to neuron 1 is removed.
Qualitatively similar results 
(Fig.~\ref{fig:HHcoupled}-\ref{fig:HHinhib}) 
are obtained with two coupled Hodgkin-Huxley neurons.
The back-action of neuron 2 on neuron 1 causes small subthreshold oscillations
and a slight decrease in the firing rate of the first neuron ($\sim$ 10\%).
These minor effects are not present in the Hodgkin-Huxley model because
the alpha function synapse does not allow back-action.  Adding buffer junctions
between the JJ neurons would be needed to obtain precise agreement.

\section{Discussion}
We have shown that the JJ Neuron is a 
biologically realistic model for single neuron dynamics, and that 
individual JJ Neurons can be coupled together in ways that are 
analogous to inhibitory or excitatory chemical synaptic coupling.  
We envision that large scale long term dynamics of networks of neurons
can be explored using JJ Neurons,
contributing to our understanding of behaviors such as
synchronization, pattern recognition, and memory formation.  
JJ Neurons should be able to alleviate computational bottlenecks
currently slowing needed simulations.
We now discuss the advantages and limitations of this approach. 

The advantages of large scale JJ Neuron simulations are speed, biological realism,
simplicity of circuit design and low power consumption.
The major advantage is speed.  
Based on similar circuits constructed for RSFQ circuits \cite{brock2001rsfq}, 
network simulations of 20,000 densely coupled neurons are reasonable and could
simulate one trillion APs for each of these neurons in a few minutes.
This speed is unachievable with digital simulations using modern computers.
Much of this advantage is due to our analog rather than digital approach.

Analog simulations using electrical circuits to model neurons 
\cite{keener1983,hoppensteadt1997introduction,douglas1998,bartolozzi2007synaptic, 
koch1998methods,cauwenberghs1999learning,hoppensteadt2006biologically}  
provide a framework more in line with the parallel nature of biological neural networks.  
Silicon-based Very-Large-Scale-Integrated (VLSI) circuits have been
designed which simulate simple 
integrate and fire neurons \cite{mead1989analog,indiveri2006}, 
bursting neurons \cite{hynna2007}, 
and plastic synapses with timing and homeostasis 
\cite{indiveri2006,bartolozzi2009}. 
The focus of VLSI research has been on implementation of a 
learning processor rather than simulation of long-term 
neuron dynamics so speed comparisons are difficult. 
Speeds of VLSI analog circuits are typically 
chosen to emulate biological neuron timescales (order milliseconds). 
Presumably these circuit speeds could be increased up to a few GHz.  
In contrast, RSFQ processors similar to the JJ Neuron 
have been clocked on the order of 100 GHz \cite{brock2001rsfq}. 
In addition, the power 
dissipation for a JJ Neuron network should be significantly 
less than a VLSI circuit of the same size, allowing larger networks.  
Finally, the non-linear behavior of Josephson junctions allows a 
JJ Neuron to contain only two Josephson junctions, much
smaller than the 22 transistors for a VLSI Integrate and Fire neuron
\cite{indiveri2006}.  

The major limitation of JJ Neuron models compared to digital simulations
is measurement.  While every neuron in a computer simulation can be
separately monitored, this is not technically feasible in large networks of 
JJ neurons.  
So for settings which require detailed monitoring of each cell in the network,
the JJ Neuron is not appropriate.
But many settings do not require such detailed observation, instead measuring 
averaged output or the output of a few key neurons.
The measurable data from JJ Neuron experiments aligns more closely with 
measurements of living tissue.  
Placing magnetic field detectors on the chip
allows measurement of averaged activity levels,  
and subsequent comparison of simulation data  
with electroencephalogram (EEG) measurements.
Other settings which involve output from only a few key neurons  
are ripe for JJ Neuron models.  
Quantities such as voltage, current or flux at important points in 
the network can be measured directly, limited by the number of wires passing
through the cooling apparatus. 
In addition, a variety of collective behaviors can be measured using 
already developed RSFQ logic circuits such as 
clocks, counters, splitters, followers, flip-flops and transmission 
lines \cite{bunyk2001rsfq}. 
So, while JJ Neuron simulations are not well suited to some questions,
they are for others, especially those involving averaged or collective 
behaviors or producing output from a small fraction of neurons. 
Analog simulations will not replace digital simulations.  
Instead, they complement each other, with analog simulations providing
collective measurement on long time scales and large networks, while
digital simulations provide detailed measurements
on shorter time scales and smaller networks.

We are pursuing a number of potential enhancements to our JJ Neuron circuit.
Along with existing tunable circuit parameters that affect the threshold,
refractory period, frequency, size and shape of APs, more flexibility
is possible by considering circuits with more junctions representing
additional ion channels.  To the extent that bursting and other exotic
neuron behavior is caused by relatively slow ion channels, 
adding a large capacitively shunted junction (with smaller characteristic
frequency) should allow JJ Neurons to display these behaviors.
Plasticity of synapses based on previous history may also be possible.  
Connection strength can be dynamically adjusted with additional circuitry
in the synapse.  If the adjustments are dependent on the firing history of the 
neurons involved, a Hebbian learning process would be possible.

\begin{acknowledgments}
We acknowledge the generous support of Colgate University's Picker 
Interdisciplinary Science Institute.  Special thanks to Bruce Hansen, Jason Myers, and
Lyle Roelofs for helpful conversations, Jon Habif, Damhnait McHugh, Terry Orlando and Joe Amato for 
helpful comments and Sarah Sciarrino for computational assistance.
\end{acknowledgments}

\bibliography{jjneuron}
\bibliographystyle{kp}

\end{document}